\def\ARXIVVERSION{1}
\def\milcomdoitext{}
\documentclass[conference]{IEEEtran}
\IEEEoverridecommandlockouts
\usepackage{balance}
\usepackage{booktabs}
\usepackage{footnote}
\usepackage{cite}
\usepackage{amsmath,amssymb,amsfonts}
\usepackage{algorithmic}
\usepackage{graphicx}
\usepackage[caption=false,font=footnotesize]{subfig}
\usepackage{textcomp}
\usepackage{xcolor}
\usepackage[normalem]{ulem}
\usepackage[nolist]{acronym}
\usepackage[all]{nowidow}
\usepackage{tabularray}
    \UseTblrLibrary{booktabs}
\usepackage[hyphens]{url}

\usepackage{tikz}
\usetikzlibrary{shapes.geometric, positioning}

\usepackage[hidelinks]{hyperref}

\newif\ifarxivversion
\ifdefined\ARXIVVERSION
    \arxivversiontrue
\else
    \arxivversionfalse
\fi
\providecommand{\milcomdoitext}{}
\providecommand{\arxivcopyrightnotice}{\textcopyright~2026 IEEE. Personal use of this material is permitted. Permission from IEEE must be obtained for all other uses, in any current or future media, including reprinting/republishing this material for advertising or promotional purposes, creating new collective works, for resale or redistribution to servers or lists, or reuse of any copyrighted component of this work in other works.}

\newif\iffinal\finalfalse 

\begin{document}

\title{SDN-Orchestrated Dual-Path 5G/SATCOM Maritime Communications for Carrier Strike Groups}

\iffinal
    \author{\IEEEauthorblockN{Anonymous Authors}}
\else
    \author{
      \IEEEauthorblockN{
        Avinash Srinivasan\IEEEauthorrefmark{1},
        Dalibor \v{S}panjevi\'{c}\IEEEauthorrefmark{2}\textsuperscript{,}\IEEEauthorrefmark{3},
        Kevin H. Nguyen\IEEEauthorrefmark{4}\textsuperscript{,}\IEEEauthorrefmark{3}, \\
        Bannon Ireton\IEEEauthorrefmark{4}\textsuperscript{,}\IEEEauthorrefmark{3}, and
        Christopher B. Landis\IEEEauthorrefmark{1}
        \thanks{\IEEEauthorrefmark{3}Work completed at the U.S. Naval Academy.}%
        \thanks{The views expressed in this document are those of the authors and do not reflect the official policy or position of the U.S. Naval Academy, Department of the Navy, the Department of War, or the U.S. Government.}
\ifarxivversion
\thanks{This is the authors' accepted manuscript of a paper accepted for publication in the 2026 IEEE Military Communications Conference (MILCOM 2026), 12--16 October 2026.\milcomdoitext\space\arxivcopyrightnotice}
\fi
      }
      \vspace{0.2cm} 
      \IEEEauthorblockA{\IEEEauthorrefmark{1}Cyber Science Department, United States Naval Academy, Annapolis, MD. Email: \{srinivas, clandis\}@usna.edu}
      \IEEEauthorblockA{\IEEEauthorrefmark{2}Carey Business School, Johns Hopkins University, Baltimore, MD. Email: dspanje1@jh.edu}
      \IEEEauthorblockA{\IEEEauthorrefmark{4}United States Navy, Pensacola, FL. Email: \{kevinnguyen2791, banto.ire20\}@gmail.com}
    }
\fi

\maketitle


\begin{acronym}[SATCOM]
\acro{AMC}{Adaptive Modulation and Coding}
\acro{BER}{bit error rate}
\acro{C2}{Command and Control}
\acro{CI}{confidence interval}
\acro{CRN}{common random numbers}
\acro{CSG}{carrier strike group}
\acro{GEO}{geostationary}
\acro{HF}{High Frequency}
\acro{IP}{Internet Protocol}
\acro{ITU}{International Telecommunication Union}
\acro{LoS}{line-of-sight}
\acro{MCS}{modulation and coding scheme}
\acro{SATCOM}{satellite communications}
\acro{SDN}{Software-Defined Networking}
\acroindefinite{SDN}{an}{a}
\acro{SDR}{Software-Defined Radio}
\acroindefinite{SDR}{an}{a}
\acro{UHF}{Ultra-High Frequency}
\acro{VHF}{Very-High Frequency}
\acro{SINR}{Signal-to-Interference-plus-Noise Ratio}
\acro{PHY}{Physical Layer}
\end{acronym}

\begin{abstract}
\ac{CSG} communications must sustain mission traffic across heterogeneous links whose quality varies with distance, weather, fading, and intermittent outages. We present an integrated \ac{SDN}/\ac{SDR} framework that unifies 5G and \ac{SATCOM} under a single control plane for dual-path failover in \ac{CSG} communications. The framework is evaluated via stochastic simulation across four operational regions: Norfolk, the Norwegian Sea, the Philippine Sea, and the North Pacific storm track. The framework couples a centralized SDN controller with a Rician-faded 5G channel under \ac{AMC}, a Ku-band \ac{SATCOM} channel with \acs{ITU}-R~P.618/P.838 rain attenuation, and Gilbert--Elliott burst-availability processes on both links. The controller routes each packet by traffic class and current link state. Experiments use 60 independent trials of 50,000 packets per parameter setting with \acl{CRN}, paired confidence intervals, and convergence diagnostics across fault, fading, and region sweeps. For the Norfolk weather baseline, SDN routing reaches 93.09\% reliability versus 36.22\% (5G-only) and 89.87\% (SATCOM-only), with paired gains of +56.87 and +3.22 percentage points. The advantage grows under joint low-Rician-$K$/high-burst-fault stress, and \ac{AMC} improves 5G robustness as \acl{SINR} degrades. SDN remains the highest-reliability mode across all four regions, indicating that cross-layer SDN-controlled failover improves maritime resilience without relying on deterministic link abstractions.
\end{abstract}

\acresetall
\section{Introduction}\label{sec:intro}

Modern \acp{CSG} operate in environments where continuous, reliable, and resilient communications are critical to mission success~\cite{maritimeCommsState-SDN-SDR}. A CSG contains an aircraft carrier, guided-missile escorts, logistics support, and other mission-dependent assets that must coordinate across wide geographic areas and dynamic threat environments. In this work, we abstract that force into a carrier with escort ships, a satellite relay, and a shore node so that routing behavior can be studied without tying the analysis to a specific country's \ac{CSG} deployment. These ships rely on a mix of 
\ac{SATCOM} links to exchange sensor data, \ac{C2} messages, and logistics information over \ac{IP}~\cite{new-era-afloat-IP}. 
However, traditional communication architectures are often hardware centric and statically configured, making it difficult to adapt to degraded \ac{LoS}, adverse weather, jamming, intermittent outages, and fluctuating bandwidth demands~\cite{maritimeCommsState-SDN-SDR,SDRarchStateChallenges}.

Conventional afloat networks also lack a unified mechanism for prioritizing traffic across heterogeneous links. A tactical air-defense update has a tighter latency and assurance requirement than routine administrative data, yet both may compete for the same physical channels and routing mechanisms. 
When rain fade, distance-dependent path loss, multipath fading, or bursty link outages degrade a path, the network may not automatically reroute traffic over alternatives such as ship--carrier--ship relays or \ac{SATCOM} backhaul. This rigidity can waste scarce satellite resources when ships are within \ac{LoS}, and it can also leave mission traffic exposed when a preferred low-latency path becomes unavailable~\cite{new-era-afloat-IP,SDRarchStateChallenges}.

\ac{SDN} and \ac{SDR} technologies offer a path toward adaptive maritime communications~\cite{SDNmeetsSDRmobileAdHocNets}.
\Iac{SDN} separates the control and data planes, enabling a logically centralized controller to observe network state, enforce mission-aware policies, and change forwarding behavior dynamically~\cite{KreutzSDN-ComprehensiveSurvey,SDNcognitiveRadioNetArch}. \Iac{SDR} complements this control-plane flexibility at the maritime radio: shipboard radios can sense sea-state-, range-, and weather-driven channel variation, adapt \ac{AMC} as link quality changes, and expose actionable link-state information to higher-layer routing decisions~\cite{SDRarchStateChallenges}.
Together, SDN and SDR motivate a dual-path cross-layer architecture in which tactical traffic normally prefers low-latency 5G, administrative traffic normally prefers \ac{SATCOM}, and the controller can fail over to avoid degraded paths. 

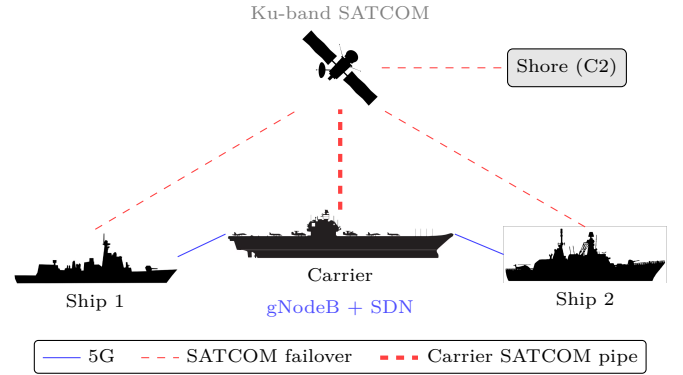
\begin{figure}[t]
\centering
\begin{tikzpicture}[
    every node/.style={font=\footnotesize},
    img/.style={inner sep=0, outer sep=0},
    shore/.style={rectangle, draw, fill=gray!20, rounded corners=2pt,
                  minimum width=11mm, minimum height=5mm, font=\scriptsize},
    fivegLink/.style={blue!75},
    satLink/.style={red!75, dashed},
    satPipe/.style={red!75, ultra thick, dashed},
]
\node[img] (sat) at (0, 2.2)
    {\includegraphics[width=1.1cm]{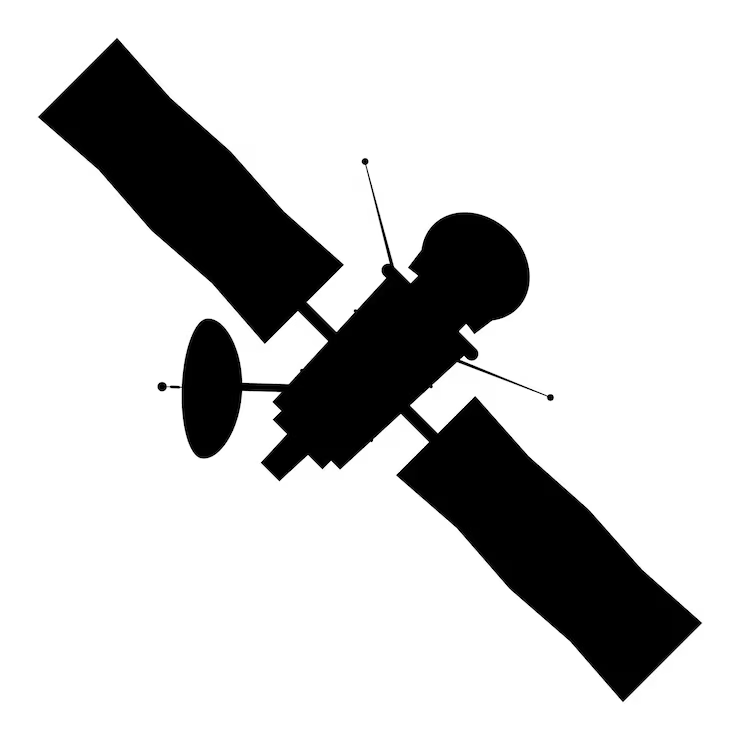}};
\node[above=-1pt of sat, font=\scriptsize, text=gray] {Ku-band SATCOM};

\node[shore] (shore) at (3.0, 2.2) {Shore (C2)};

\node[img] (carrier) at (0, 0)
    {\includegraphics[scale=0.10]{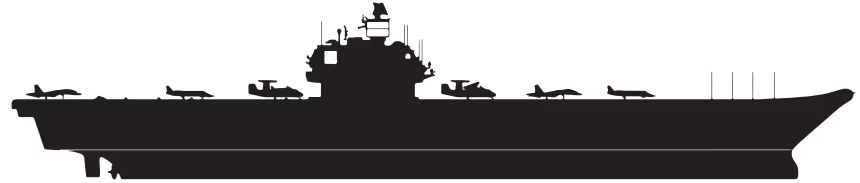}};
\node[below=0pt of carrier, font=\scriptsize] (carrierLabel) {Carrier};
\node[below=0pt of carrierLabel, font=\scriptsize, text=blue!70] {gNodeB + SDN};

\node[img] (s1) at (-3.24, -0.3)
    {\includegraphics[scale=0.20]{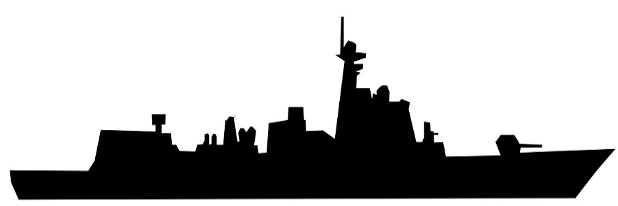}};
\node[below=0pt of s1, font=\scriptsize] {Ship 1};

\node[img] (s2) at (3.24, -0.27)
    {\includegraphics[scale=0.15]{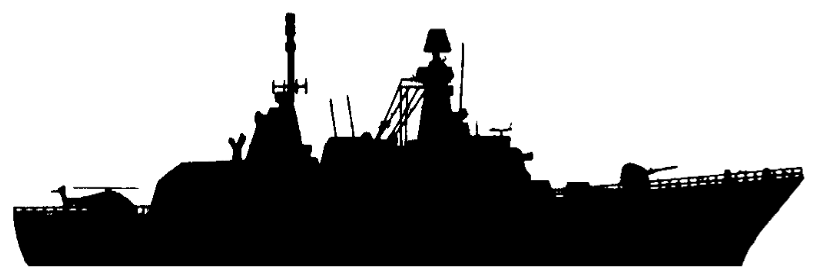}};
\node[below=0pt of s2, font=\scriptsize] {Ship 2};

\draw[fivegLink] (carrier.west) -- (s1.east);
\draw[fivegLink] (carrier.east) -- (s2.west);

\draw[satLink] (s1.north) -- (sat.south west);
\draw[satLink] (s2.north) -- (sat.south east);

\draw[satPipe] (carrier.north) -- (sat.south);

\draw[satLink] (sat.east) -- (shore.west);

\node[draw, fill=white, anchor=north, font=\scriptsize,
      rounded corners=1.5pt, inner sep=3pt] at (0, -1.4) {%
\tikz[baseline=-0.5ex] \draw[fivegLink] (0,0)--(5mm,0);~5G \quad
\tikz[baseline=-0.5ex] \draw[satLink]   (0,0)--(5mm,0);~SATCOM failover \quad
\tikz[baseline=-0.5ex] \draw[satPipe]   (0,0)--(5mm,0);~Carrier SATCOM pipe
};
\end{tikzpicture}
\vspace{-2mm}
\caption{Simulated \ac{CSG}: carrier-hosted gNodeB/SDN controller, escorts with 5G primary, SATCOM failover, \ac{GEO} ship--shore relay, and shore node. Two of five escorts shown.}
\label{CSG-topology}
\vspace{-2mm}
\end{figure}

Programmable maritime networks combining 5G and \ac{SATCOM} under unified \ac{SDN}/\ac{SDR} control have been proposed in prior architectural work~\cite{maritimeCommsState-SDN-SDR,SDNmeetsSDRmobileAdHocNets,SDNcognitiveRadioNetArch}. However, many simulation studies typically rely on deterministic link abstractions or scenario-level fault gates that do not capture packet-level variability in fading, weather, and burst errors~\cite{devStratifiedApproachSDNsimulation,SDN4CoRE}. This paper addresses these gaps by implementing a physically grounded stochastic simulation: a Rician-faded 5G link with \ac{AMC}, a Ku-band \ac{SATCOM} link with \ac{ITU}-R~P.618~\cite{ITU-R-P618} rain attenuation, and Gilbert--Elliott burst-error availability for both paths, evaluated under paired stochastic conditions.
The simulated topology (\figurename~\ref{CSG-topology}) comprises the aircraft carrier (hosting gNodeB and SDN controller), five escort ships, a \ac{GEO} satellite for ship-shore relay, and a shore command node; details are formalized in Section~\ref{sec:topology}. The controller classifies each packet as tactical or administrative, observes link state and \ac{PHY} feasibility, and selects between the 5G and \ac{SATCOM} paths with per-packet failover. Performance is measured using packet delivery reliability, tactical deadline-miss rate, \ac{SATCOM} offload, latency jitter, and mean latency. The key contributions of this work are:
\begin{itemize}
    \item A stochastic dual-path \ac{CSG} simulator that captures packet-level variability in fading, weather, and link availability across both 5G and \ac{SATCOM} paths, enabling realistic operational assessment under stochastic link stress.
    \item A per-packet mission-aware SDN controller with cross-layer use of channel state, AES-256-GCM accounting for tactical traffic on \ac{SATCOM}, and three routing modes: \texttt{sdn}, \texttt{static\_5g} (5G-only), and \texttt{static\_satcom} (SATCOM-only) for paired comparison.
    \item A rigorous evaluation protocol providing statistically grounded inter-mode comparisons via paired-sample inference and convergence diagnostics.
    \item A multi-region evaluation showing how fault rate, Rician $K$-factor, traffic mix, and weather distributions affect SDN reliability and failover behavior.
\end{itemize}

The remainder of this paper is organized as follows. Section~\ref{sec:rw} reviews related work on SDN, SDR, hybrid maritime links, and resilient network simulation. Section~\ref{sec:method} presents the system model and experimental methodology. Section~\ref{sec:results} reports the simulation results, sensitivity studies, and multi-region validation. Section~\ref{sec:conclusion} concludes the paper with future research directions.

\section{Related Work}\label{sec:rw}

Prior maritime SDN/SDR work has built the architectural case for programmable control over heterogeneous shipboard links~\cite{SDNmeetsSDRmobileAdHocNets,SDNcognitiveRadioNetArch,GongSurveySDN-Applications}.
In general, a logically centralized controller can manage paths, allocate scarce spectrum, and react to changing link conditions in ways that statically configured naval \ac{IP} infrastructure cannot~\cite{new-era-afloat-IP}. Cross-layer designs that expose \ac{SDR} channel state to higher-layer routing decisions further extend this control surface to the physical layer~\cite{SDRarchStateChallenges,SDRplatformsWirelessTech}. Niknami et al.~\cite{maritimeCommsState-SDN-SDR} proposed a unified SDN--SDR-driven cross-layer maritime communications framework that leverages the existing \ac{SATCOM} infrastructure for resilient communications in dynamic, resource-constrained environments; our presented work implements and quantitatively evaluates that framework under stochastic channel, weather, and burst-availability conditions. These prior contributions are largely architectural; they motivate \ac{SDN} control for naval communications but stop short of such evaluation.

\subsection{SDN for Multi-SATCOM and Cognitive Maritime Links}
A complementary thread targets the multi-SATCOM aggregation problem. The SDN-SAT lineage models each ship as an SDN switch over several segregated \ac{SATCOM} terminals and uses Multipath TCP under centralized control for load balancing and throughput optimization~\cite{nazari2016software,du2017traffic}, with later extensions adding UAV relays for handover continuity in mobile tactical scenarios~\cite{zhao2018software}. Ghafoor and Koo~\cite{ghafoor2020cognitive} take a different route, layering SDN-based cognitive routing over cognitive-radio maritime links with cluster-head local views. These efforts validate their controllers in Mininet-class emulators or custom simulators with opaque links, leaving the packet-level interaction between \ac{PHY}-layer impairments and \ac{SDN} routing decisions outside their scope.

\subsection{Simulation and Fault Modeling}
\ac{SDN} simulation studies for resilient networks typically validate controller logic, topology effects, and failover behavior under deterministic outages, independent periodic faults, or scenario-level weather gates~\cite{devStratifiedApproachSDNsimulation,SDN4CoRE}. Such abstractions can obscure packet-level effects that drive maritime link behavior, including multipath fading, weather-driven attenuation, and bursty outages. \Ac{CRN} and paired statistical comparisons separate routing-policy effects from random environmental variation~\cite{LawSimulation}, but this discipline is rarely combined with physically grounded \ac{PHY} models in the maritime \ac{SDN} literature.

\section{System Model and Methodology}\label{sec:method}

This section presents the simulation model: a physically-grounded, stochastic, dual-path link model coupled to a centralized SDN controller with an adaptive SDR front end. Relative to the deterministic abstractions common in prior maritime SDN studies~\cite{devStratifiedApproachSDNsimulation,SDNmeetsSDRmobileAdHocNets}, the model comprises:
\begin{enumerate}
    \item a Rician-faded 5G channel with \ac{AMC};
    \item a Ku-band SATCOM channel with \ac{ITU}-R~P.618 rain attenuation driven by per-quarter-hour observed weather;
    \item a Gilbert--Elliott~\cite{gilbert1960capacity,elliott1963estimates} two-state link-availability model; and
    \item an experimental protocol with \ac{CRN}, paired hypothesis tests, and run-count and packet-count convergence diagnostics following Law and Kelton~\cite{LawSimulation}.
\end{enumerate}

\subsection{Topology, Traffic, and Operational Regions}\label{sec:topology}
The \ac{CSG} is modeled as seven surface nodes: one aircraft carrier (co-located gNodeB and SDN controller host), five guided-missile escorts at a mean baseline carrier-to-escort separation of $\approx 20$~km (ranging $1.1$ to $44$~km across the high-, mid-, and low-band 5G envelopes), and one shore command facility. Surface node positions are specified by geodesic coordinates and pairwise inter-node separations are computed via the Haversine formula,
\vspace{-2mm}
\begin{multline}
d =2R_E \arcsin\\\left(\sqrt{\sin^2\!\left(\frac{\Delta\phi}{2}\right)
   + \cos\phi_1 \cos\phi_2 \sin^2\!\left(\frac{\Delta\lambda}{2}\right)}\right),
\label{eq:haversine}
\end{multline}
where $R_E = 6{,}371$~km is Earth's mean radius, $\phi$ is latitude, and $\lambda$ is longitude. The \ac{GEO} \ac{SATCOM} relay is treated as a separate link abstraction characterized by a fixed elevation angle ($45^\circ$) and an effective rain-layer path length, rather than as a positioned node in the topology. The SATCOM channel is modeled with a fixed slant-range geometry, capturing rain fade, propagation delay, and return-channel overhead. A per-region treatment with elevation-angle-dependent rain attenuation is left to future work.

To isolate the effect of regional weather distributions, the same CSG topology and SATCOM geometry are simulated under weather samples drawn from four operational regions: Norfolk, VA, USA (mid-Atlantic baseline), the Norwegian Sea ($68^\circ$N, polar edge), the Philippine Sea (tropical), and the North Pacific storm track ($50^\circ$N, $-155^\circ$W). Differences in metrics across regions thus reflect differences in rain-rate distributions rather than fleet geometry.

Traffic is generated as a stream of packets denoted by $N_p$, each tagged \emph{tactical} or \emph{administrative}. Tactical packets are short (256\,B nominal), latency-sensitive with a representative deadline $D_\text{tac}=100$~ms (chosen as a low-latency tactical messaging target), and prefer the lowest-latency feasible link; administrative packets are larger (4\,kB) and tolerant of SATCOM transit. The fraction of generated packets tagged tactical, $\pi_\text{tac}$, is a sweep parameter (default $0.5$, equal mix).

\subsection{Dual-Path PHY Model}

\subsubsection{5G Link with Rician Fading and \texorpdfstring{\ac{AMC}}{AMC}}
For each packet, the instantaneous 5G \ac{SINR}, $\gamma=S/(I+N)$~\cite{TseWireless2005}, is drawn from a Rician distribution parameterized by $K$-factor, where $K$ controls the ratio of deterministic \ac{LoS} power to diffuse multipath power, and a deterministic path-loss term whose breakpoints determine band selection: high-band ($d\!\le\!1.6$~km), mid-band ($1.6\!<\!d\!\le\!20$~km), and low-band ($20\!<\!d\!\le\!55$~km), where $55$~km is the 5G coverage cutoff beyond which packets are routed to SATCOM. Conditioned on \ac{SINR} $\gamma$, the radio selects the highest \ac{MCS} whose minimum \ac{SINR} is met (Table~\ref{tab:amc})~\cite{GoldsmithChua1997,TseWireless2005}. This makes the SDR element of the framework explicit: the \ac{PHY} senses the channel and adapts. An ablation with \ac{AMC} disabled fixes the scheme and drops packets whose \ac{SINR} cannot support it.

\begin{table}[t]
\centering
\caption{\ac{AMC} table used by the 5G radio.}
\label{tab:amc}
\vspace{-2mm}
\begin{tabular}{lrr}
\toprule
MCS & Min SINR (dB) & Spectral Eff.\ (b/s/Hz) \\
\midrule
BPSK     & $2$   & 0.5 \\
QPSK     & $5$   & 1.0 \\
16-QAM   & $10$  & 2.0 \\
64-QAM   & $16$  & 4.0 \\
256-QAM  & $22$  & 6.0 \\
\bottomrule
\end{tabular}
\vspace{-2mm}
\end{table}

\subsubsection{Ku-band SATCOM with Rain Fade}
The SATCOM channel models propagation delay, transmission delay, and Ku-band rain attenuation per Recommendation \ac{ITU}-R~P.618~\cite{ITU-R-P618}. For specific attenuation $\gamma_R$ (dB/km) at frequency $f$ and rain rate $R$ (mm/h), with frequency-dependent coefficients $k(f),\alpha(f)$ tabulated in \ac{ITU}-R~P.838~\cite{ITU-R-P838},
\vspace{-1mm}
\begin{equation}
\gamma_R = k(f)\,R^{\alpha(f)},
\label{eq:itu-gammaR}
\vspace{-2mm}
\end{equation}
with slant-path attenuation $A = \gamma_R\,L_E(\theta,h_R)$ accumulated along an effective path length $L_E$ that depends on elevation $\theta$ and rain-cell height $h_R$. The end-to-end SATCOM latency is\vspace{-1mm}
\begin{equation}
\tau_\text{sat} = \frac{2nh}{c} + \frac{L}{R_b} + n_r\,T_r,
\label{eq:tau-sat}
\vspace{-2mm}
\end{equation}
where $n$ counts hops through the \ac{GEO} relay and shore node, $L$ is payload size, $R_b\in\{2,4,8\}$~Mbps is the negotiated data rate, $n_r$ is the retry count, and $T_r$ a retry penalty.

\subsubsection{Weather Sampling}
Rain rate is treated as a per-packet stochastic input rather than a scenario-level constant. The simulator ingests one year (May 2025--April 2026) of 15-min Open-Meteo observations at the chosen operational region, and for each packet draws an independent and identically distributed (i.i.d.) rain-rate sample $R$ from this empirical distribution. The sampled $R$ enters the SATCOM link budget as \ac{ITU}-R~P.618 rain attenuation, so fade events affect individual packets according to the observed weather distribution of the region rather than its annual mean. The i.i.d.\ per-packet sampling is a deliberate worst-case stress test of the routing layer: it preserves the marginal rain-rate distribution observed in the region but breaks the temporal correlation of sustained storm episodes, exposing the controller to maximum-variance rain-fade transitions on every packet. Coupling rain rate to the Gilbert--Elliott bad-state process to model storm persistence as a continuous-time Markov channel is left as future work; the current results should be read as an upper bound on per-packet fade variability under realistic marginal weather.

\subsection{Link Availability: Gilbert--Elliott Model}
Independent of fading and rain, each link transitions between \emph{good} and \emph{bad} states as a two-state Markov chain~\cite{gilbert1960capacity,elliott1963estimates}. While bad, the link drops packets at a high error probability; while good, error probability is set by the \ac{PHY}. The chain is parameterized by the steady-state bad probability $\pi_B$ and the mean bad-state burst length $\bar{L}_B$ (Table~\ref{tab:ge}); these jointly fix the transition matrix.

\begin{table}[t]
\centering
\caption{Gilbert--Elliott link-availability parameters (baseline).}
\label{tab:ge}
\vspace{-2mm}
\begin{tabular}{lrr}
\toprule
Parameter & 5G & SATCOM \\
\midrule
Steady-state bad prob.\ $\pi_B$       & 0.05  & 0.02  \\
Mean bad-burst length $\bar{L}_B$ (packets) & 6    & 8    \\
Error prob.\ in good state          & (PHY)  & (PHY)  \\
Error prob.\ in bad state           & 1.00  & 1.00  \\
\bottomrule
\end{tabular}
\vspace{-2mm}
\end{table}

\subsection{SDN Control Plane and Routing}
The SDN controller is co-located with the carrier and maintains a global view of node geometry, per-link Gilbert--Elliott state, the \ac{AMC}-selected rate on each 5G link, and the SATCOM availability. The routing policy is per-packet and class-aware. The four elements below describe primary-path selection, failover, encryption of tactical traffic on SATCOM, and the routing modes used for paired comparison.
\begin{enumerate}
    \item \textbf{Primary path.} Tactical packets prefer 5G if (a)~the destination is within band range, (b)~the link is in the good state, and (c)~\ac{SINR} supports at least the lowest \ac{MCS}. Administrative packets prefer SATCOM unless the rain-fade margin is exceeded.
    \item \textbf{Failover.} On primary-path unavailability, the controller routes to the alternate path, incurring a configurable controller processing delay $T_c$.
    \item \textbf{Encryption.} AES-256-GCM is applied to tactical traffic crossing SATCOM; the latency contribution $T_e$ is recorded separately for the decomposition analysis.
    \item \textbf{Routing Modes.} Three modes share the same simulator for paired comparison: \texttt{sdn} (dual-path with failover), \texttt{static\_5g}, and \texttt{static\_satcom}.
\end{enumerate}
\vspace{-2mm}
\subsection{Performance Metrics}
Each independent trial reports: \emph{Reliability}, the fraction of packets delivered; \emph{Tactical deadline-miss rate} ($D_\text{tac}$), the fraction of tactical packets exceeding; \emph{SATCOM offload}, the fraction of delivered packets carried on the satellite path; \emph{Latency jitter}, defined as $p_{99}\!-\!p_{50}$ of end-to-end latency on the 5G path; and \emph{Mean latency}. For one detail-logged independent trial, latency is decomposed into five contributions $\tau = \tau_\text{prop}+\tau_\text{tx}+\tau_\text{retry}+\tau_\text{ctrl}+\tau_\text{enc}$.

\subsection{Experimental and Statistical Methodology}
The simulation follows discrete-event simulation practice from Law and Kelton~\cite{LawSimulation}, using \ac{CRN} for paired comparison, convergence-validated sample sizes, and parameter sweeps for sensitivity and multi-region analysis.

\subsubsection{CRN} For each independent-trial index $r\!\in\!\{1,\dots,R\}$, all modes draw fading, rain, Gilbert--Elliott transitions, and traffic from a seeded stream identified by $r$. The three modes therefore see paired environments, which sharpens SDN-vs-static contrast and is required for paired t-test described below.

\subsubsection{Paired t-test} Let $X^{(\textsc{sdn})}_r$ and $X^{(\textsc{base})}_r$ denote a metric on independent trial $r$ under SDN and a static baseline. We test $H_0\!:\!\mathbb{E}[X^{(\textsc{sdn})}-X^{(\textsc{base})}]=0$ via one-sample $t$ statistic on paired differences $D_r$, reporting the mean difference, 95\% \ac{CI} $\bar{D}\pm t_{0.975,R-1}\,s_D/\sqrt{R}$, and $t$ statistic.

\subsubsection{Sample sizes and convergence} Each parameter setting uses $R\!=\!60$ independent trials of $N_p\!=\!50{,}000$ packets, with both choices justified by convergence sweeps on the Reliability metric. Sub-sampling $R\in\{30,40,50,60\}$ from the $R\!=\!60$ experiment yields 95\% \ac{CI} half-widths on Reliability that drop below $0.1~\mathrm{pp}$ at $R\!=\!60$. A separate sweep with $N_p\in\{20{,}000,\,30{,}000,\,40{,}000,\,50{,}000\}$ confirms that the mean reliability estimate moves by less than $0.1~\mathrm{pp}$ across this range, well within the per-run Monte Carlo \ac{CI}.

\subsubsection{Sensitivity and joint sweeps} Univariate sweeps over Rician $K$-factor (covering the range observed in over-water maritime channel measurements~\cite{wang2018wireless}), ship spacing, and tactical traffic share~$\pi_\text{tac}$ are run with the same $R\!=\!60$ independent-trial budget and reported with 95\% \ac{CI} bands. A two-dimensional sweep over the Gilbert--Elliott bad-state probability and the Rician $K$-factor characterizes the joint operating region in which the SDN dual-path advantage is largest.

\subsubsection{Multi-region study} The baseline configuration is rerun under each of the four regions defined in Section~\ref{sec:topology}, with the per-region weather record as the only varying input.

\section{Results}\label{sec:results}

The configurations detailed in Section~\ref{sec:method} led to the following results.
The per-mode metrics are summarized in Table~\ref{tab:results-summary}; the paired SDN-vs-static comparisons are in Table~\ref{tab:paired-compare}.

\begin{table}[t]
\centering
\caption{Per-mode reliability and latency (Norfolk baseline).\\ SatOff (SATCOM offload) shown only for SDN\\ (trivially 0\%/100\% for static modes).}
\vspace{-2mm}
\label{tab:results-summary}
\begin{tblr}{colspec = {lrrrr}, row{1} = {m}, cell{1}{2-Z} = {c}, rowsep = 0.3pt}
\toprule
Mode & {Reliability\\(\%)} & {TDMiss\\(\%)} & {SatOff\\(\%)} & {Mean Latency\\(ms)} \\
\midrule
\texttt{sdn}            & 93.09 & 73.97 & 19.77 & 515.29 \\
\texttt{static\_5g}     & 36.22 & 73.93 & ---   & 124.07 \\
\texttt{static\_satcom} & 89.87 & 100   & ---   & 612.37 \\
\bottomrule
\end{tblr}
\vspace{-2mm}
\end{table}

\begin{table}[t]
\centering
\caption{Paired SDN-vs-static comparison on Reliability\\ ($R=60$, percentage points).}
\label{tab:paired-compare}
\vspace{-2mm}
\begin{tabular}{lrrr}
\toprule
Baseline & $\overline{\Delta}$ (pp) & 95\% CI (pp) & $t$ \\
\midrule
\texttt{static\_5g}     & 56.87 & $[56.79, 56.95]$ & 1416.88 \\
\texttt{static\_satcom} & 3.22  & $[3.12, 3.31]$   & 68.52   \\
\bottomrule
\end{tabular}
\vspace{-2mm}
\end{table}

\subsection{Routing-Mode Comparison}
\begin{figure}[t]
\centering
\subfloat[Per-trial reliability by routing mode ($R\!=\!60$).\label{fig:reliability-by-mode}]{%
    \includegraphics[width=0.51\columnwidth]{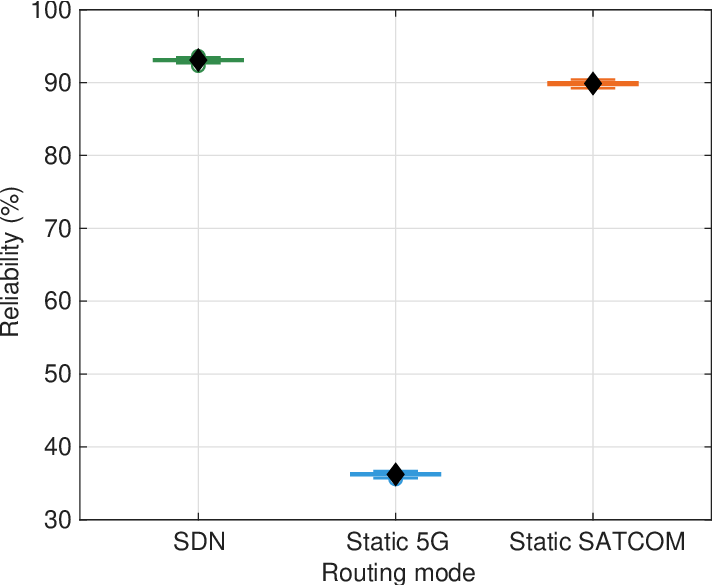}}
\hfill
\subfloat[Reliability vs.\ Gilbert--Elliott $\pi_B$ fault probability with 95\% CI bands.\label{fig:fault-sweep}]{%
    \includegraphics[width=0.47\columnwidth]{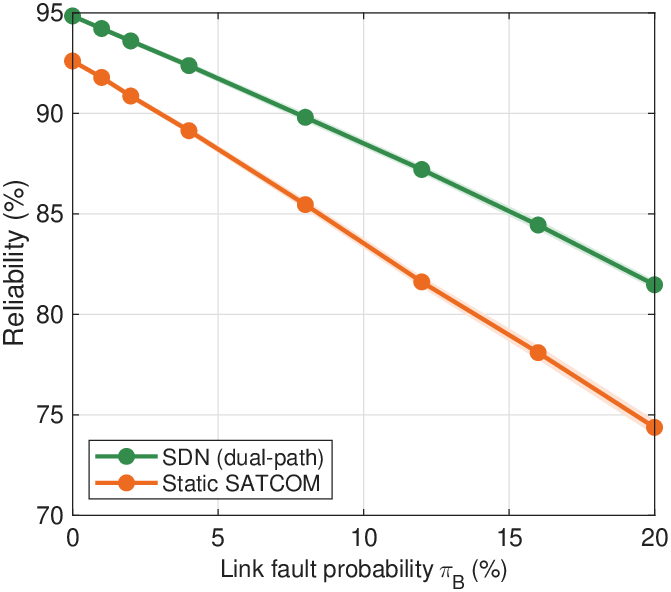}}
\vspace{-1mm}
\caption{Reliability}
\label{fig:reliability-summary}
\vspace{-2mm}
\end{figure}

SDN achieves $93.09\%$ mean reliability versus $36.22\%$ 5G-only and $89.87\%$ SATCOM-only (Table~\ref{tab:results-summary}). Paired improvements are $+56.87~\mathrm{pp}$ over \texttt{static\_5g} (95\% CI $[56.79,56.95]$) and $+3.22~\mathrm{pp}$ over \texttt{static\_satcom} (95\% CI $[3.12,3.31]$, Table~\ref{tab:paired-compare}). The modest $+3.22~\mathrm{pp}$ gain over SATCOM-only understates the operational picture: SATCOM-only misses every tactical deadline ($\text{TDMiss}=100\%$), so its delivery rate is irrelevant for tactical traffic.
But SDN's failover preserves a tactical-capable path SATCOM alone cannot. The marginal increase in tactical deadline-miss rate (\texttt{sdn} $73.97\%$ vs.\ \texttt{static\_5g} $73.93\%$) reflects this trade explicitly: SDN failover routes $19.8\%$ of tactical traffic to SATCOM, where the \ac{GEO} transit time alone ($\approx$480~ms) exceeds the $100$~ms deadline. The framework thus trades a sub-percentage-point increase in deadline-miss for a $56.87$~pp gain in delivery reliability, a worthwhile trade when packet loss is operationally costlier than latency, and one that motivates lower-orbit satellite integration to recover the latency margin. 
\figurename~\ref{fig:reliability-by-mode} shows SDN's per-run reliability above both static baselines across all 60 runs, with no overlap between modes.

Mean latency (Table~\ref{tab:results-summary}) for 5G-only is inverse: the lowest mean latency ($124$~ms) is \texttt{static\_5g} because failures do not affect latency but delivery rate. \texttt{static\_satcom} has $612$~ms mean latency with and no packet meeting the $100$~ms tactical deadline. SDN is between the two ($515$~ms mean) and, as previously noted, offloads $19.8\%$ of delivered traffic to SATCOM, keeping reliability above $93\%$. SATCOM latency is dominated by the \ac{GEO} propagation term; the SDN-controller processing delay $T_c$ and AES-256-GCM overhead $T_e$ together contribute sub-millisecond latency per packet, negligible relative to the \ac{GEO} transit time on the SATCOM path.

\subsection{Behavior Under Varying Link Faults}
Sweeping the Gilbert--Elliott bad-state probability $\pi_B$ from light to severe degradation, \figurename~\ref{fig:fault-sweep} shows SDN reliability remaining above the \texttt{static\_satcom} baseline across the swept range, with 95\% \ac{CI} bands that do not overlap the operating range. The static modes provide visual lower and upper reliability envelopes without path agility. The SDN controller's failover activates progressively as $\pi_B$ rises, with SATCOM's share of delivered traffic growing monotonically across the swept range. Total delivery loss is correspondingly mitigated relative to the \texttt{static\_5g} baseline.


\subsubsection{AMC Impact on 5G Performance}
\begin{figure}[t]
\centering
\subfloat[Tactical deadline-miss rate.\label{fig:amc-deadline}]{%
    \includegraphics[width=0.49\columnwidth]{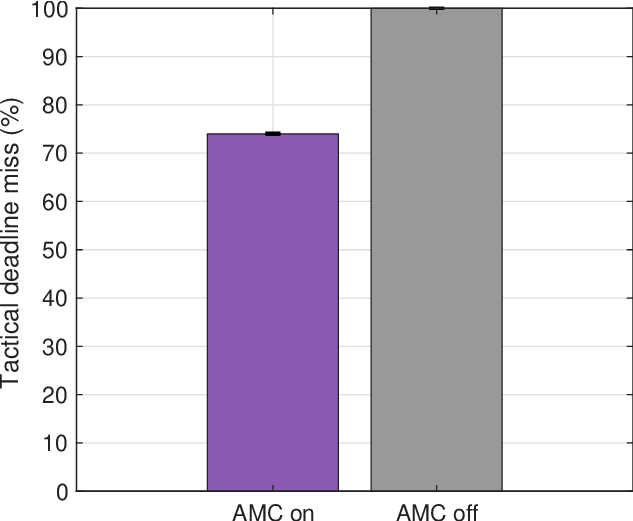}}
\hfill
\subfloat[5G Throughput.\label{fig:amc-throughput}]{%
    \includegraphics[width=0.49\columnwidth]{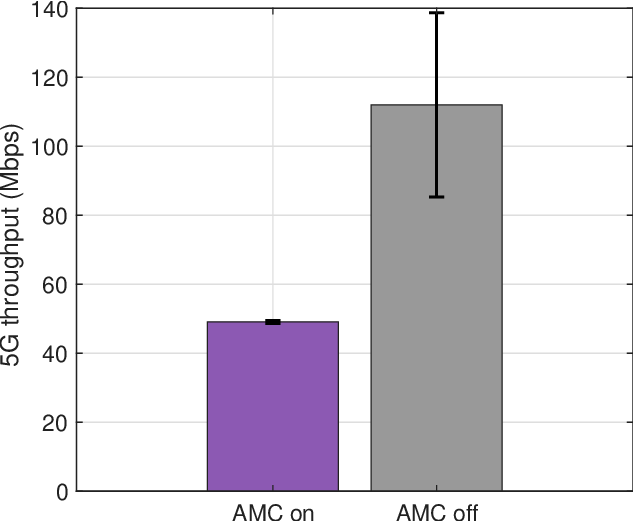}}
\vspace{-1mm}
\caption{\ac{AMC} trades peak throughput for robustness as \ac{SINR} falls.}
\label{fig:amc}
\vspace{-2mm}
\end{figure}

The \ac{AMC} ablation isolates the SDR contribution. The operational benefit is in \figurename~\ref{fig:amc-deadline}: with \ac{AMC} disabled, the fixed high-\ac{MCS} scheme misses the tactical deadline on every packet ($\text{TDMiss}=100\%$), so the 5G path is unviable for tactical traffic.
Enabling \ac{AMC} drops the radio to a lower \ac{MCS} rather than dropping the packet when \ac{SINR} falls, cutting TDMiss to $74\%$. The throughput trade is in \figurename~\ref{fig:amc-throughput}: \ac{AMC} produces a tightly-clustered $\approx 49$~Mbps mean rate, while the fixed-\ac{MCS} configuration shows a higher mean ($\approx 112$~Mbps) but only on its successful subset of packets and with substantially larger variance. The trade-off is clear: \ac{AMC} sacrifices peak per-packet rate for predictability and tactical viability.

\begin{table*}[ht]
\centering
\caption{Per-region SDN metrics; weather sampled from a year of 15-min observations in each region.}
\label{tab:location-compare}
\vspace{-2mm}
\begin{tabular}{lcccccc}
\toprule
Location & Reliability (\%) & TacDeadlineMiss (\%) & SatOff (\%) & Jitter5G (ms) & MeanLatency (ms) & PeakRain (mm/hr) \\
\midrule
Norfolk & 93.09 & 73.97 & 19.77 & 375.77 & 515.29 & 77.70 \\
NorwegianSea & 89.11 & 74.22 & 20.68 & 375.66 & 506.59 & 91.80 \\
PhilippineSea & 91.29 & 74.40 & 19.37 & 375.76 & 514.42 & 58.20 \\
NorthPacific & 88.42 & 74.29 & 20.77 & 375.64 & 504.96 & 87.00 \\
\bottomrule
\end{tabular}
\vspace{-2mm}
\end{table*}

\subsubsection{Formation Geometry and 5G Path Utilization}
\begin{figure}[t]
\centering
\vspace{-2mm}
\subfloat[Rician $K$-factor sensitivity.\label{fig:spacing-k}]{%
    \includegraphics[width=0.48\columnwidth]{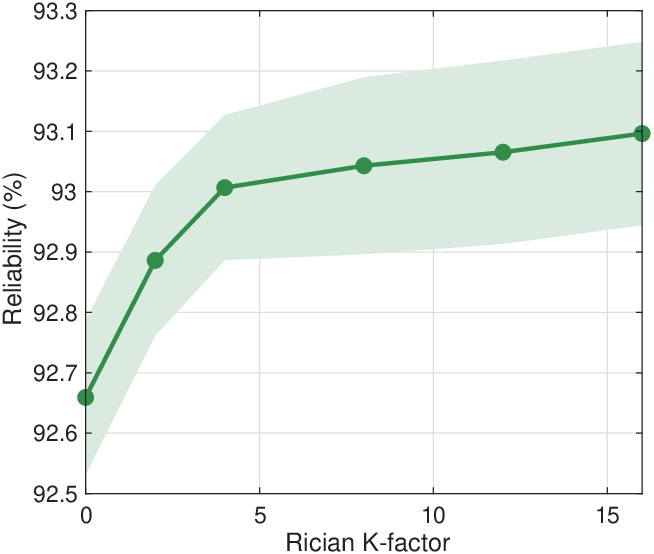}}
\hfill
\subfloat[Ship-spacing sensitivity.\label{fig:spacing-distance}]{%
    \includegraphics[width=0.48\columnwidth]{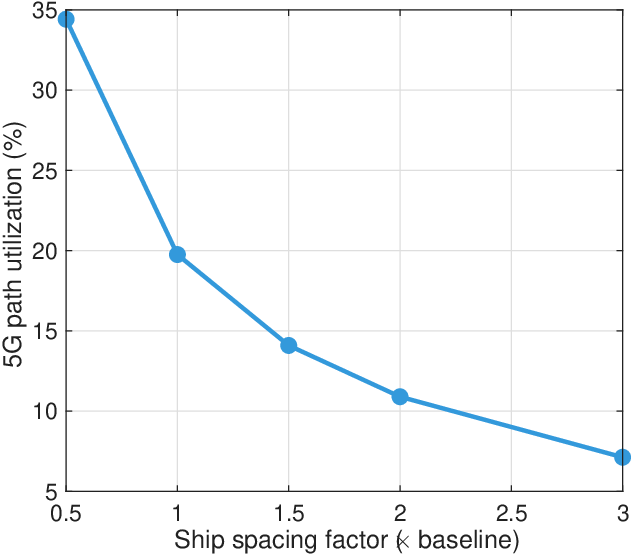}}
    \vspace{-1mm}
\caption{SDN sensitivity to fading and formation geometry.}
\label{fig:spacing}
\vspace{-2mm}
\end{figure}

CSG formations are not fixed: screen distances vary with mission phase, sea state, and threat posture. \figurename~\ref{fig:spacing-k} shows SDN reliability is robust to Rician $K$-factor, varying under $0.5$pp across $K\in[0,16]$ with overlapping confidence bands; the key result is not tied to a specific channel-quality assumption. \figurename~\ref{fig:spacing-distance} sweeps inter-ship spacing multiplicatively over baseline formation: 5G path utilization declines monotonically from $\approx 35\%$ at $0.5\times$ to under $10\%$ at $3\times$. As ships disperse, fewer pairs remain within high- and mid-band 5G range envelopes, and the controller correctly migrates traffic to SATCOM. 

\subsubsection{Rain-Fade Effects on SATCOM}
The per-packet rain-rate sampling and \ac{ITU}-R~P.618 attenuation model~\cite{ITU-R-P618} produce \ac{BER} fluctuations consistent with the predicted slant-path attenuation curve, with \ac{BER} rising sharply above the rain-fade margin (\figurename~\ref{fig:ber-rain}). The corresponding hour-of-day reliability dip tracks the empirical rain-rate distribution at each region; reliability is otherwise stable across the diurnal cycle.

\subsubsection{Joint Operating Region}
\begin{figure}[t]
\centering
\subfloat[\label{fig:ber-rain}]{%
    \includegraphics[width=0.475\columnwidth]{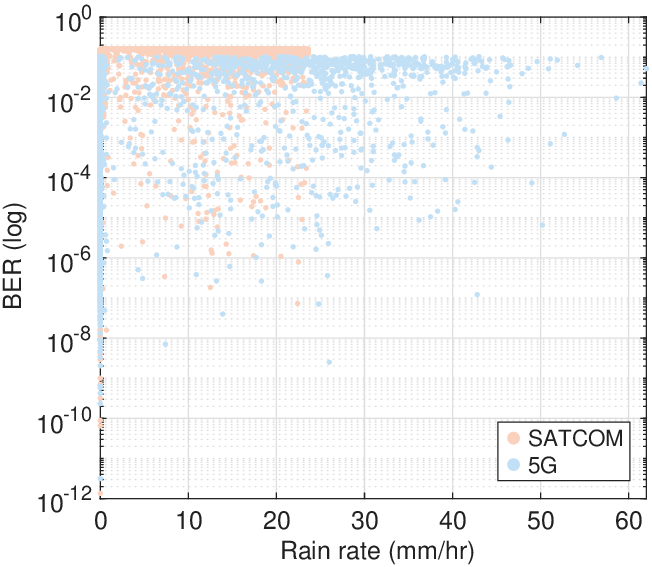}%
}\hfil
\subfloat[\label{fig:joint-gain}]{%
    \includegraphics[width=0.5\columnwidth]{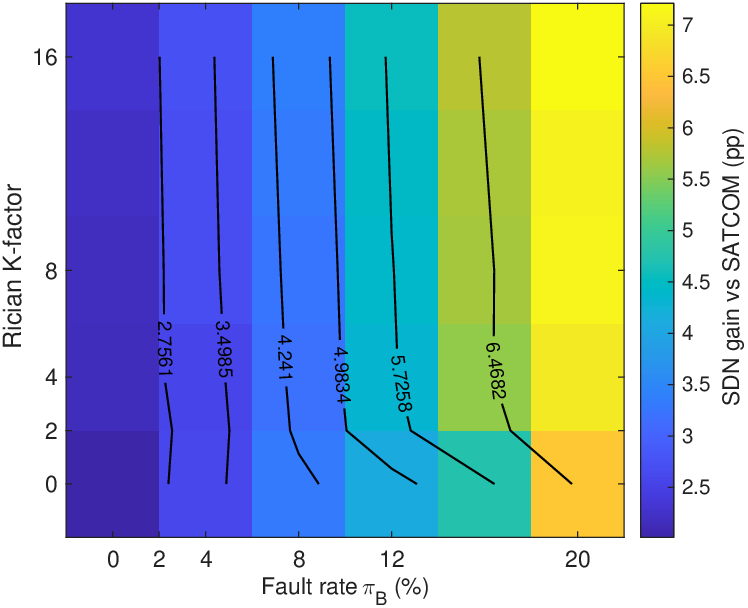}%
}
\caption{Physical-layer validation and system-level characterization: (a) \ac{BER} vs.\ rain rate on the Ku-band SATCOM link with the \ac{ITU}-R~P.618/P.838 prediction overlaid; (b) SDN reliability gain across the joint $(\pi_B, K)$ parameter space with overlaid gain contours.}
\label{fig:characterization}
\vspace{-2mm}
\end{figure}

To characterize where the SDN dual-path advantage is largest, \figurename~\ref{fig:joint-gain} plots the reliability gain of SDN over the better static baseline as a function of both $\pi_B$ and Rician $K$. The advantage is largest in the joint stress region (low $K$, high $\pi_B$) and small but positive in clear-sky conditions, which is the operationally relevant pattern: SDN matters most where conditions are worst on both axes.


\subsubsection{Multi-Region Validation}
 Holding the \ac{CSG} topology fixed and varying only the per-region weather record (Table~\ref{tab:location-compare}), SDN reliability ranges from $88.4\%$ (North Pacific storm track) to $93.1\%$ (Norfolk), with the Norwegian Sea polar-edge region at $89.1\%$ and the Philippine Sea at $91.3\%$. The differences track the per-region peak rain rate more closely than they track latitude. The framework holds across all four regions, and SDN remains the highest-reliability mode in each operational region, CI widths remaining comparable.


\section{Conclusion}\label{sec:conclusion}
We presented a stochastic dual-path SDN/SDR simulator for \ac{CSG} communications, coupling Rician 5G with \ac{AMC} and Ku-band SATCOM with \ac{ITU}-R~P.618 rain fade under a Gilbert--Elliott availability model. Paired comparison against \texttt{static\_5g} and \texttt{static\_satcom} baselines shows that SDN dual-path routing yields significantly higher reliability while sustaining the $100$~ms tactical deadline that \texttt{static\_satcom} cannot meet, with the \ac{AMC} element contributing quantifiable robustness as \ac{SINR} falls. The framework holds across four maritime regions, including a high-latitude polar-edge region.
Future work includes adversarial jamming and contested-spectrum scenarios, snow attenuation at high latitudes, lower-orbit satellite integration to recover tactical-latency margins on the failover path, high-fidelity SATCOM geometry computed from \ac{GEO} sub-satellite longitudes with elevation-angle-dependent rain fade (capturing the elevation degradation at regions such as the Norwegian Sea and North Pacific), time-correlated rain-rate sampling to model storm persistence, 5G PC5 sidelink for direct ship-to-ship tactical paths, and waveform-level SDR modeling in Simulink.

\balance
\bibliographystyle{IEEEtran}
\bibliography{refs}

\end{document}